\documentclass[10pt, conference, compsocconf]{sp2017_template/IEEEtran}
\ifCLASSINFOpdf
\else
\fi

\usepackage[utf8]{inputenc} 
\usepackage[T1]{fontenc}    
\usepackage{url}            
\usepackage{booktabs}       
\usepackage{amsfonts}       
\usepackage{nicefrac}       
\usepackage{microtype}      
\usepackage{amsmath} 
\usepackage{algorithm}
\usepackage{algorithmic}
\usepackage{graphicx}
\usepackage{subcaption}
\usepackage{authblk}
\usepackage{color, soul}
\usepackage[draft,inline,nomargin]{fixme}

\def\sys{FakeGAN}

\def\l2lambda{0.2}

\def\accnosim{89.1\%}

\def\stateacc{$\sim 89.8\%$}
\def\dprimeacc{$\sim 99\%$}

\newcommand{\citep}{\cite}
\newcommand{\citet}{\cite}

\title{Detecting Deceptive Reviews using Generative Adversarial Networks}

\author[1]{Hojjat Aghakhani\thanks{hojjat@cs.ucsb.edu}}
\author[1]{Aravind Machiry\thanks{B.B@university.edu}}
\author[2]{Shirin Nilizadeh\thanks{C.C@university.edu}}
\author[1]{Christopher Kruegel\thanks{D.D@university.edu}}
\author[1]{Giovanni Vigna\thanks{E.E@university.edu}}
\affil[1]{University of California, Santa Barbara}
\affil[1]{\{hojjat, machiry, chris, vigna\}@cs.ucsb.edu}
\affil[2]{Carnegie Mellon University Silicon Valley}
\affil[2]{shirin.nilizadeh@sv.cmu.edu}

\begin{document}

\maketitle

\begin{abstract}
In the past few years, consumer review sites have become the main target of \emph{deceptive opinion spam}, where fictitious opinions or reviews are deliberately written to sound authentic.
Most of the existing work to detect the deceptive reviews focus on building supervised classifiers based on syntactic and lexical patterns of an opinion. With the successful use of Neural Networks on various classification applications, in this paper, we propose \sys{} a system that for the first time augments and adopts Generative Adversarial Networks (GANs) for a text classification task, in particular, detecting deceptive reviews. 

Unlike standard GAN models which have a single Generator and Discriminator model, \sys{} uses two discriminator models and one generative model. The generator is modeled as a stochastic policy agent in reinforcement learning (RL), and the discriminators use Monte Carlo search algorithm to estimate and pass the intermediate action-value as the RL reward to the generator. 
Providing the generator model with two discriminator models avoids the mod collapse issue by learning from both distributions of truthful and deceptive reviews.
Indeed, our experiments show that using two discriminators provides \sys{} high stability, which is a known issue for GAN architectures. 
While \sys{} is built upon a \emph{semi-supervised classifier}, known for less accuracy, our evaluation results on a dataset of TripAdvisor hotel reviews show the same performance in terms of accuracy as of the state-of-the-art approaches that apply supervised machine learning. 
These results indicate that GANs can be effective for text classification tasks. Specifically, \sys{} is effective at detecting deceptive reviews.
\end{abstract}

\section{Introduction}

In the current world, we habitually turn to the wisdom of our peers, and often complete strangers, for advice, instead of merely taking the word of an advertiser or business owner. A 2015 study by marketing research company Mintel~\citep{Mintel2015} found nearly 70 percent of Americans seek out others' opinions online before making a purchase. Many platforms such as Yelp.com and TripAdvisor.com have sprung up to facilitate this sharing of ideas amongst users. 
The heavy reliance on review information by the users has dramatic effects on business owners. It has been shown that an extra half-star rating on Yelp helps restaurants to sell out 19 percentage points more frequently~\citep{anderson2012learning}.

\begin{figure*}[t!]
    \centering
    \begin{subfigure}[t]{0.5\textwidth}
        \centering
        \includegraphics[height=1.2in]{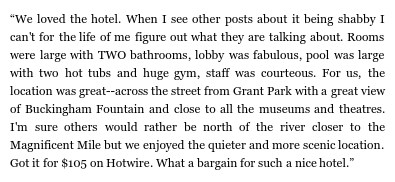}
        \caption{A truthful review provided by a high profile user on TripAdvisor}
    \end{subfigure}%
    ~ 
    \begin{subfigure}[t]{0.5\textwidth}
        \centering
        \includegraphics[height=1.2in]{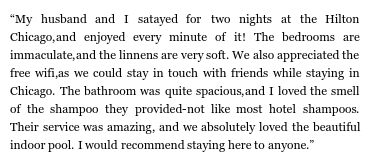}
        \caption{A deceptive review written by an Amazon Mechanical worker}
    \end{subfigure}
    \caption{A truthful review versus a deceptive review, both written for the same hotel.}
    \label{fig:example}
\end{figure*}

This phenomenon has also lead to a market for various kinds of fraud. In simple cases, this could be a business rewarding its customers with a discount, or outright paying them, to write a favorable review.  In more complex cases, this could involve astroturfing, opinion spamming~\citep{Jindal:2008} or \emph{deceptive opinion spamming}~\cite{ott2011finding}, where fictitious reviews are deliberately written to sound authentic. Figure~\ref{fig:example} shows an example of a truthful and deceptive review written for the same hotel. It is estimated that up to 25\% of Yelp reviews are fraudulent~\citep{yelp-fake-per, luca2016fake}.

Detecting deceptive reviews is a text classification problem. In recent years, deep learning techniques based on natural language processing have been shown to be successful for text classification tasks. 
 Recursive Neural Network (RecursiveNN)~\citet{socher2011semi, socher2011dynamic, socher2013recursive} has shown good performance classifying texts, while 
Recurrent Neural Network (RecurrentNN)~\citet{elman1990finding} 
 better captures the contextual information and is ideal for realizing semantics of long texts. 
 However, RecurrentNN is a biased model, where later words in a text have more influence than earlier words~\citet{lai2015recurrent}. This is not suitable for tasks such as detection of deceptive reviews that depend on an unbiased semantics of the entire document (review).
Recently, techniques based on Convolutional Neural Network (CNN)~\citet{kim2014convolutional, zhang2015character} were shown to be effective for text classification. However, the effectiveness of these techniques depends on careful selection of the window size~\citep{lai2015recurrent}, which controls the parameter space.

Moreover, in general, the main problem with applying classification methods for detecting deceptive reviews is the lack of substantial ground truth datasets required for most of the supervised machine learning techniques. This problem worsens for neural networks based methods, whose complexity requires much bigger dataset to reach a reasonable performance. 

To address the limitations of the existing techniques, we propose \sys{}, which is a technique based on Generative Adversarial Network (GAN)~\citet{goodfellow2014generative}. 
GANs are a class of artificial intelligence algorithms used in unsupervised machine learning, implemented by a system of two neural networks contesting with each other in a zero-sum game framework. 
GANs have been used mostly for image-based applications~\cite{goodfellow2014generative, radford2015unsupervised, ehsani2017segan, denton2015deep}.
In this paper, for the first time, we propose the use of GANs for a text classification task, i.e., detecting deceptive reviews. 
Moreover, the use of a semi-supervised learning method like GAN can eliminate the problem of ground truth scarcity that in general hinders the detection success~\citet{ott2011finding, li2014towards, yoo2009comparison}.

We augment GAN models for our application in such a way that unlike standard GAN models which have a single Generator and Discriminator model, \sys{} uses two discriminator models $D$, $D'$ and one generative model $G$.
The discriminator model $D$ tries to distinguish between truthful and deceptive reviews whereas $D'$ tries to distinguish between reviews generated by the generative model $G$ and samples from \emph{deceptive} reviews distribution.
The discriminator model $D'$ helps $G$ to generate reviews close to the deceptive reviews distribution, while $D$ helps $G$ to generate reviews which are classified by $D$ as truthful. 

Our intuition behind using two discriminators is to create a stronger generator model. If in the adversarial learning phase, the generator gets rewards only from $D$, the GAN may face the mod collapse issue~\citet{metz2016unrolled}, as it tries to learn two different distributions (truthful and deceptive reviews). 
The combination of $D$ and $D'$ trains $G$ to generate better deceptive reviews which in turn train $D$ to be a better discriminator. 

Indeed, our evaluation using the TripAdvisor\footnote{Tripadvisor.com} hotel reviews dataset shows that the discriminator $D$ generated by \sys{} performs on par with the state-of-the-art methods that apply supervised machine learning, with an accuracy of~\accnosim{}. 
These results indicate that GANs can be effective for text classification tasks, specifically, FakeGAN is effective at detecting deceptive reviews. To the best of our knowledge, \sys{} is the first work that use GAN to generate better discriminator model (i.e., $D$) in contrast to the common GAN applications which aim to improve the generator model.

In summary, following are our contributions:
\begin{enumerate}
\item We propose \sys{}, a deceptive review detection system based on a double discriminator GAN.
\item We believe that~\sys{} demonstrates a good first step towards using GANs for text classification tasks. 
\item To the best of our knowledge, \sys{} is the first system using semi-supervised neural network-based learning methods for detecting deceptive fraudulent reviews.
\item Our evaluation results demonstrate that \sys{} is as effective as the state-of-the-art methods that apply supervised machine learning for detecting deceptive reviews. 
\end{enumerate}

\section{Approach} \label{approach}


Generative Adversarial Network (GAN)~\citet{goodfellow2014generative} is a promising framework for generating high-quality samples with the same distribution as the target dataset. 
\sys{} leverages GAN to learn the distributions of truthful and deceptive reviews and to build a semi-supervised classifier using the corresponding distributions. 

A GAN consists of two models: a generative model $G$ which tries to capture the data distribution, and a discriminative model $D$ that distinguishes between samples coming from the training data or the generator $G$. These two models are trained simultaneously, where $G$ is trying to fool the discriminator $D$, while $D$ is maximizing its probability estimation that whether a sample comes from the training data or is produced by the generator. In a nutshell, this framework corresponds to a minimax two-player game. 

The feedback or the gradient update from discriminator model plays a vital role in the effectiveness of a GAN. In the case of text generation, it is difficult to pass the gradient update because the generative model produces discrete tokens (words), but the discriminative model makes a decision for complete sequence or sentence. 
Inspired by SeqGAN~\citet{yu2017seqgan} that uses GAN model for Chinese poem generation, in this work, we model the generator as a stochastic policy in reinforcement learning (RL), where the gradient update or RL reward signal is provided by the discriminator using Monte Carlo search. 
Monte Carlo is a heuristic search algorithm for identifying the most promising moves in a game. In summary, in each state of the game, it plays out the game to the very end for a fixed number of times according to a given policy. To find the most promising move, it must be provided by reward signals for a complete sequence of moves.

All the existing applications use GAN to create a strong generator, where the main issue is the convergence of generator model~\citet{arjovsky2017wasserstein, gulrajani2017improved, metz2016unrolled}. 
\textit{Mode collapse} in particular is a known problem in GANs, where complexity and multimodality of the input distribution cause the generator to produce samples from a single mode. The generator may switch between modes during the learning phase, and this cat-and-mouse game may never end~\citet{goodfellow2016nips, metz2016unrolled}. 
Although no formal proof exists for convergence, in Section~\ref{sec:exper} we show that the \sys{}'s discriminator converges in practice.

Unlike the typical applications of GANs, where the ultimate goal is to have a strong generator, \sys{} leverages GAN to create a well-trained discriminator, so that it can successfully distinguish truthful and deceptive reviews.
However, to avoid the stability issues inherent to GANs we augment our network to have two discriminator models though we use only one of them as our intended classifier. Note that leveraging samples generated by the generator makes our classifier a \emph{semi-supervised} classifier.


\begin{figure}[h]
    \centering
    \includegraphics[width=0.45\textwidth]{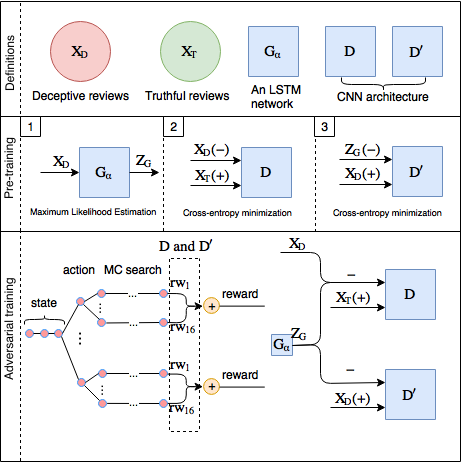}
    \caption{The overview of \sys{}. The symbols $+$ and $-$ indicates positive and negative samples respectively. Note that, these are different from truthful and deceptive reviews.}
    \label{fig:fakegan}
\end{figure}
\subsection*{Definitions}
We start with defining certain symbols which will be used throughout this section to define various steps of our approach. The training dataset, $X = X_D \cup X_T $, consists of two parts, deceptive reviews $X_D$ and truthful reviews $X_T$.  We use $\chi$ to denote the vocabulary of all tokens (i.e., words) which are available in $X$. 

Our generator model $G_{\alpha}$ parametrized by $\alpha$ produces each review $S_{1:L}$ as a sequence of tokens of length $L$ where $S_{1:L} \in \chi^{L}$.  We use $Z_G$ to indicate all the reviews generated by our generator model $G_{\alpha}$.

We use two discriminator models $D$ and $D'$. The discriminator $D$ distinguishes between truthful and deceptive reviews, as such $D(S_{1:L})$ is the probability that the sequence of tokens comes from $X_T$ or $X_D \cup Z_G$. Similarly, $D'$ distinguishes between deceptive samples in the dataset and samples generated by $G_{\alpha}$ consequently $D'(S_{1:L})$ is a probability indicating how likely the sequence of tokens comes from $X_D$ or $Z_G$. 

The discriminator $D'$ guides the generator $G_{\alpha}$ to produce samples similar to $X_D$ whereas $D$ guides $G_{\alpha}$ to generate samples which seems truthful to $D$. So in each round of training,  by using the feedback from $D$ and $D'$, the generator $G_{\alpha}$ tries to fool $D'$ and $D$ by generating reviews that seems deceptive (not generated by $G_{\alpha}$) to $D'$, and truthful (not generated by $G_{\alpha}$ or comes from $X_D$) to $D$. 

Figure~\ref{fig:fakegan} shows an overview of \sys{}. During pre-training, we use the Maximum Likelihood Estimation (MLE) to train the generator $G_{\alpha}$ on deceptive reviews $X_D$ from the training dataset. We also use minimizing the cross-entropy technique to pre-train the discriminators. 

The generator $G_{\alpha}$ is defined as a policy model in reinforcement learning. In timestep $t$, the state $s$ is the sequence of produced tokens, and the action $a$ is the next token. The policy model $G_{\alpha}(S_t|S_{1:t-1})$ is stochastic. Furthermore, the generator $G_{\alpha}$ is trained by using a policy gradient and Monte Carlo (MC) search on the expected end reward from the discriminative models $D$ and $D'$.  Similar to~\cite{yu2017seqgan}, we consider the estimated probability $D(S_{1:L}) + D'(S_{1:L})$ as the reward. Formally, the corresponding action-value function is:
\begin{equation} \label{eq:reward}
A_{G_{\alpha}, D, D'}(a=S_L, s=S_{1:L-1}) = D(S_{1:L}) + D'(S_{1:L})
\end{equation}

As mentioned before, $G_{\alpha}$ produces a review token by token. However, the discriminators provide the reward for a complete sequence. Moreover, $G_{\alpha}$ should care about the long-term reward, similar to playing Chess where players sometimes prefer to give up immediate good moves for a long-term goal of victory~\citep{silver2016mastering}. Therefore, to estimate the action-value function in every timestep $t$, we apply the Monte Carlo search $N$ times with a roll-out policy $G'_{\gamma}$ to sample the undetermined last $L-t$ tokens. We define an $N$-time Monte Carlo search as
\begin{equation} \label{eq:MC}
    \{S_{1:L}^1, S_{1:L}^2, ..., S_{1:L}^N\}=MC_{G'_{\gamma}}(S_{1:t}, N)
\end{equation}
where for $1 \leq i \leq N$
\begin{equation}
    S_{1:t}^i = (S_1, ..., S_t)
\end{equation}
and $S_{t+1:L}^i$ is sampled via roll-out policy $G'_{\gamma}$ based on the current state $S_{1:t-1}^i$. The complexity of action-value estimation function mainly depends on the roll-out policy. While one might use a simple version (e.g., random sampling or sampling based on n-gram features) as the policy to train the GAN fast, to be more efficient, we use the same generative model ($G'_{\gamma}=G_{\alpha}$ at time $t$). Note that, a higher value of $N$ results in less variance and more accurate evaluation of the action-value function. We can now define the action-value estimation function at $t$ as
\begin{multline} \label{eq:estimation}
A_{G_{\alpha}, D, D'}(a=S_t, s=S_{1:t-1})= \\
\begin{cases}
    \frac{1}{N} \sum_{i=1}^N (D(S_{1:L}^i) + D'(S_{1:L}^i)) & \text{if } t \leq L\\
    D(S_{1:L}) + D'(S_{1:L})              & \text{if } t = L
\end{cases}
\end{multline}
where $S_{1:L}^i$s are created according to the Equation~\ref{eq:MC}. 
As there is no intermediate reward for the generator, we define the the objective function for the generator $G_{\alpha}$ (based on \citet{sutton2000policy}) to produce a sequence from the start state $S_0$ to maximize its final reward:
\begin{equation} \label{eq:objectiveFunction}
J(\alpha) = \sum_{S_1 \in \chi}G_{\alpha}(S_1|S_0) \text{ }. \text{ } A_{G_{\alpha}, D, D'}(a=S_1, s=S_0)
\end{equation}
Conseqently, the gradient of the objective function $J(\alpha)$ is:
\begin{equation} \label{eq:objectiveFuncUpdate}
\resizebox{0.9\hsize}{!}{$\displaystyle\nabla _{\alpha} J(\alpha) = \sum_{t=1}^{T}\mathbb{E}_{S_{1:t-1}\sim G_{\alpha}}[\sum_{S_t \in \chi} \nabla _{\alpha} G_{\alpha}(S_t|S_{1:t-1}) \text{ } . \text{ } A_{G_{\alpha}, D, D'}(a=S_t, s=S_{1:t-1})]$}
\end{equation}
We update the generator's parameters ($\alpha$) as:
\begin{equation} \label{eq:policyGradUpdate}
\alpha \leftarrow \alpha + \lambda \nabla _{\alpha} J(\alpha)
\end{equation}
where $\lambda$ is the learning rate.

By dynamically updating the discriminative models, we can further improve the generator. So, after generating $g$ samples, we will re-train the discriminative models $D$ and $D'$ for $d$ steps using the following objective functions respectively:
\begin{equation} \label{eq:discUptade}
min(-\mathbb{E}_{S \sim X_T} [\log D(S)] - \mathbb{E}_{S \sim X_D \vee G_{\alpha}} [1 - \log D(S)])
\end{equation}
\begin{equation} \label{eq:discUptade2}
min(-\mathbb{E}_{S \sim X_D} [\log D'(S)] - \mathbb{E}_{S \sim G_{\alpha}} [1 - \log D'(S)])
\end{equation}
In each of the $d$ steps, we use $G_{\alpha}$ to generate the same number of samples as number of truthful reviews i.e., $|X_{G}| = |X_{T}|$.
The updated discriminators will be used to update the generator, and this cycle continues until \sys{} converges. Algorithm~\ref{alg1} formally defines all the above steps.


\begin{algorithm}
\caption{\sys{}}
\label{alg1}
\begin{algorithmic}
    \REQUIRE discriminators $D$ and $D'$, generator $G_{\alpha}$, roll-out policy $G_{\gamma}$, dataset $X$
    \STATE Initialize $\alpha$ with random weight.
    \STATE Load word2vec vector embeddings into $G_{\alpha}$, $D$ and $D'$ models
    \STATE Pre-train $G_{\alpha}$ using MLE on $X_D$
    \STATE Pre-train $D$ by minimizing the cross entropy
    \STATE Generate negative examples by $G_{\alpha}$ for training $D'$
    \STATE Pre-train $D'$ by minimizing the cross entropy
    \STATE $\gamma \leftarrow \alpha$
    \REPEAT
        \FOR{g-steps}
            \STATE Generate a sequence of tokens $S_{1:L} = (S_1, ..., S_L) \sim G_{\alpha}$
            \FOR{$t$ in $1:L$}
                \STATE Compute $A_{G_{\alpha}, D_{\beta}, D'_{\theta}}(a=S_t, s=S_{1:t-1})$ by Eq. \ref{eq:estimation}
            \ENDFOR
            \STATE Update $\alpha$ via policy gradient Eq. \ref{eq:policyGradUpdate}
        \ENDFOR
        \FOR{d-steps}
            \STATE Use $G_{\alpha}$ to generate $X_{G}$.
            \STATE Train discriminator $D$ by Eq. \ref{eq:discUptade}
            \STATE Train discriminator $D'$ by Eq. \ref{eq:discUptade2}
        \ENDFOR
        \STATE $\gamma \leftarrow \alpha$
    \UNTIL{D reaches a stable accuracy.}
\end{algorithmic}
\end{algorithm}

\subsection*{The Generative Model}
We use RecurrentNNs (RNNs) to construct the generator. An RNN maps the input embedding representations $s_1, ..., s_L$ of the input sequence of tokens $S_1, ..., S_L$ into hidden states $h_1, ..., h_L$ by using the following recursive function. 
\begin{equation}
h_t = g(h_{t-1}, s_t)
\end{equation}
Finally, a softmax output layer $z$ with bias vector $c$ and weight matrix $V$ maps the hidden layer neurons into the output token distribution as
\begin{equation}
p(s|s_1, ..., s_t) = z(h_t) = \text{softmax}(c+V . h_t)
\end{equation}
To deal with the common vanishing and exploding gradient problem \citep{goodfellow2016deep} of the backpropagation through time, we exploit the Long Short-Term Memory (LSTM) cells \citep{hochreiter1997long}.
\subsection*{The Discriminator Model}
For the discriminators, we select the CNN because of their effectiveness for text classification tasks~\citep{zhang2015text}. First, we construct the matrix of the sequence by concatenating the input embedding representations of the sequence of tokens $s_1, ..., s_L$ as:
\begin{equation}
\zeta_{1:L}=s_1\oplus...\oplus s_L
\end{equation}
Then a kernel $w$ computes a convolutional operation to a window size of $l$ by using a non-linear function $\pi$, which results in a feature map:
\begin{equation}
f_i=\pi(w\otimes \zeta_{i:i+l-1} + b)
\end{equation}
Where $\otimes$ is the inner product of two vectors, and $b$ is a bias term. Usually, various numbers of kernels with different window sizes are used in CNN. 
We hyper-tune size of kernels by trying kernels which have been successfully used in text classification tasks by community~\citet{zhang2015character, wang2017liar, lai2015recurrent}. Then we apply a max-over-time pooling operation over the feature maps to allow us to combine the outputs of different kernels. Based on \citep{srivastava2015highway} we add the highway architecture to improve the performance. In the end, a fully connected layer with sigmoid activation functions is used to output the class probability of the input sequence.

\section{Evaluation} \label{sec:exper}
We implemented \sys{} using the TensorFlow~\cite{abadi2016tensorflow} framework. We chose the dataset from~\citep{ott2011finding} which has 800 reviews of 20 Chicago hotels with positive sentiment. The dataset consists of 400 truthful reviews provided by high profile users on TripAdvisor and 400 deceptive reviews written by Amazon Mechanical Workers. 
To the best of our knowledge, this is the biggest available dataset of labeled reviews and has been used by many related works~\citet{ott2011finding, li2014towards, feng2012syntactic}. 
Similar to SeqGAN~\citet{yu2017seqgan}, the generator in \sys{} only creates fixed length sentences.
Since the majority of reviews in this dataset has a length less than 200 words, we set the sequence length of \sys{} ($L$) to 200. For sentences whose length is less than 200, we pad them with a fixed token <END> to reach the size of 200 resulting in 332 truthful and 353 deceptive reviews. Note that, having a larger dataset results in a less training time. Although larger dataset makes each adversarial step slower, it provides $G$ a richer distribution of samples, thus reduces the number of adversarial steps resulting in less training time.

We used the k-fold cross-validation with k=5 to evaluate \sys{}. We leveraged GloVe vectors\footnote{Check ``glove.6B.200d.txt'' from https://nlp.stanford.edu/projects/glove/} for word representation~\citep{pennington2014glove}. Similar to SeqGAN~\citet{yu2017seqgan}, the convergence of \sys{} varies with the training parameters $g$ and $d$ of generator and discriminative models respectively. After experimenting with different values, we observed that following values $g = 1$ and $d = 6$ are optimal.
For pre-training phase, we trained the generator and the discriminators until convergence, which took 120 and 50 steps respectively. The adversarial learning starts after the pre-training phase. All our experiments were run on a 40-core machine, where the pre-training took $\sim$one hour, and the adversarial training took $\sim$11 hours with a total of $\sim$12 hours. 

\subsection{Accuracy of Discriminator $D$}
\begin{figure*}[t!]
    \centering
    \begin{subfigure}[t]{0.45\textwidth}
        \centering
        \includegraphics[width=0.75\textwidth]{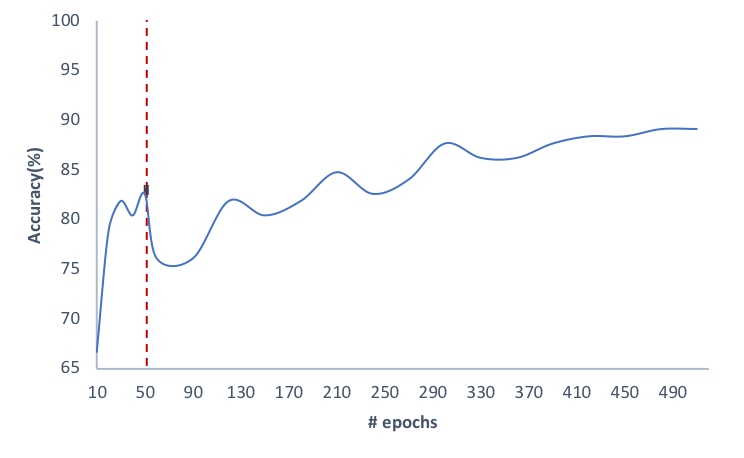}
    \caption{Accuracy of \sys{} (Discriminator $D$) at each step by feeding the testing dataset to $D$. While minimizing cross entropy method for pre-training $D$ converges and reaches accuracy at $\sim 82\%$, adversarial training phase boosts the accuracy to $\sim 89\%$.}
    \label{fig:learningd}
    \end{subfigure}%
    ~ 
    \begin{subfigure}[t]{0.45\textwidth}
        \centering
        \includegraphics[width=0.75\textwidth]{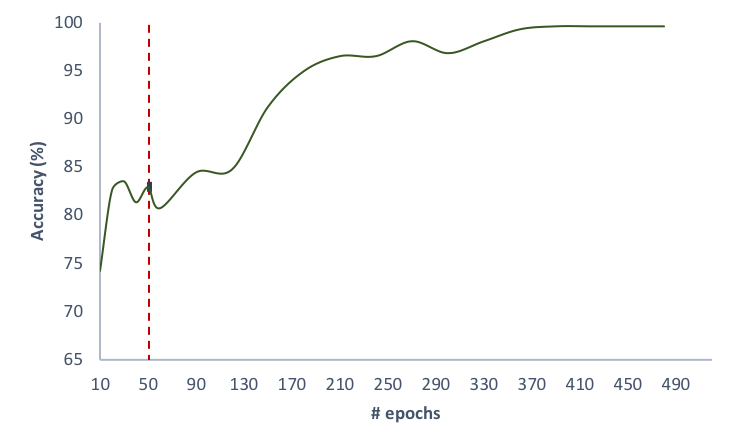}
    \caption{Accuracy of $D'$ at each step by feeding the testing dataset and generated samples by $G$ to $D'$. Similar to figure~\ref{fig:learningd}, this plot shows that $D'$ converged after 450 steps resulting in the convergence of \sys{}.}
    \label{fig:learningdprime}
    \end{subfigure}
    \caption{The accuracy of $D$ and $D'$ on the test dataset over epochs. The vertical dashed line shows the beginning of adversarial training.}
\end{figure*}

As mentioned before, the goal of \sys{} is to generate a highly accurate discriminator model, $D$, that can distinguish deceptive and truthful reviews. Figure~\ref{fig:learningd} shows the accuracy trend for this model; for simplicity, the trend is shown only for the first iteration of k-fold cross-validation.
During the pre-training phase, the accuracy of $D$ stabilized at $50^{th}$ step. We set the adversarial learning to begin at step 51. After a little decrease in accuracy at the beginning, the accuracy increases and converges to $89.2\%$, which is on-par with the accuracy of state-of-the-art approach~\citet{ott2011finding} that applied supervised machine learning on the same dataset (\stateacc{}). The accuracy, precision and recall for k-fold cross-validation are 89.1\%, $98\%$ and $81\%$ all with a standard deviation of 0.5. 
This supports our hypothesis that adversarial training can be used for detecting deceptive reviews. Interestingly even though \sys{} relies on semi-supervised learning, it yields similar performance as of a fully-supervised classification algorithm.

\subsection{Accuracy of Discriminator $D'$}
Figure~\ref{fig:learningdprime} shows the accuracy trend for the discriminator $D'$. Similar to $D$, $D'$ converges after 450 steps with an accuracy of \dprimeacc{} accuracy. It means that at this point, the generator $G$ will not be able to make any progress trying to fool $D'$, and the output distribution of $G$ will stay almost same. Thus, continuing adversarial learning does not result in any improvement of the accuracy of our main discriminator, $D$.

\subsection{Comparing \sys{} with the original GAN approach}
To justify the use of two discriminators in \sys{}, we tried using just one discriminator (only $D$) in two different settings. In the first case, the generator $G$ is pre-trained to learn only \emph{truthful reviews} distribution. Here the discriminator $D$ reached $83\%$ accuracy in pre-training, and the accuracy of adversarial learning, i.e., the classifier, reduces to about $65\%$. 
In the second case, the generator $G$ is pre-trained to learn only \emph{deceptive reviews} distribution. Unlike the first case, adversarial learning improved the performance of $D$ by converging at $84\%$, however, still, the performance is lower than that of \sys{}. 

These results demonstrate that using two discriminators is necessary to improve the accuracy of \sys{}. 

\subsection{Scalability Discussion}
We argue that the time complexity of our proposed augmented GAN with two discriminators is the same as of original GANs because their bottleneck is the MC search, where using the rollout policy (which is $G$ until the time) generates 16 complete sequences, to help the generator $G$ for just outputting the most promising token as its current action. This happens for every token of a sequence which is generated by $G$. However, compared to MC search, discriminators $D$ and $D'$ are efficient and not time-consuming.

\subsection{Stability Discussion}
\begin{figure*}[t!]
    \centering
    \begin{subfigure}[t]{0.45\textwidth}
        \centering
        \includegraphics[width=0.75\textwidth]{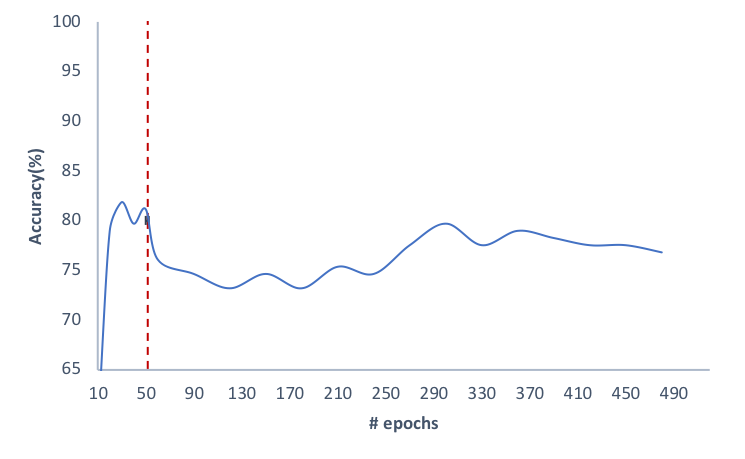}
    \caption{The accuracy of $D$ fluctuates around 77\% in constrast to the stabilization at $89.1\%$ in Figure~\ref{fig:learningd} (with values g=1 and d=6)}
    \end{subfigure}%
    ~ 
    \begin{subfigure}[t]{0.45\textwidth}
        \centering
        \includegraphics[width=0.75\textwidth]{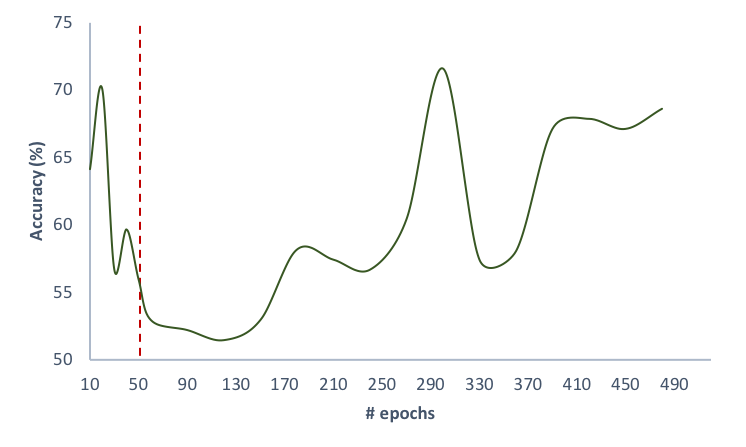}
    \caption{Accuracy of $D'$. Unlike in Figure~\ref{fig:learningdprime}, this plot shows that $D'$ is not stable.}
    \end{subfigure}
    \caption{The accuracy of $D$ and $D'$ on the test dataset over epochs while both $g$ and $d$ are one. 
    }
    \label{learning-dg1}
\end{figure*}

As we discussed in Section~\ref{approach}, the \emph{stability} of GANs is a known issue. We observed that the parameters $g$ and $d$ have a large effect on the convergence and performance of \sys{} as illustrated in the Figure~\ref{learning-dg1}, when $d$ and $g$ are both equal to one. We believe that the stability of GAN makes hyper-tuning of \sys{} a challenging task thus prevents it from outperforming the state-of-the-art methods based on supervised machine learning. However, with the following values $d=6$ and $g=1$, \sys{} converges and performs on par with the state-of-the-art approach.

\section{Related work}
%
Text classification has been used extensively in email spam~\citep{drucker1999support} detection and link spam detection in web pages~\citep{gyongyi2004combating, ntoulas2006detecting, gyongyi2005web}.  Over the last decade, researchers have been working on \emph{deceptive opinion spam}.

Jindal et al.~\citet{Jindal:2008} first introduced \emph{deceptive opinion spam} problem as a widespread phenomenon and showed that it is different from other traditional spam activities. They built their ground truth dataset by considering the duplicate reviews as spam reviews and the rest as nonspam reviews. They extracted features related to review, product and reviewer, and trained a Logistic Regression model on these features to find fraudulent reviews on Amazon. Wu et al.~\citet{wu2010distortion} claimed that deleting dishonest reviews will distort the popularity significantly. They leveraged this idea to detect deceptive opinion spam in the absence of ground truth data. Both of these heuristic evaluation approaches are not necessarily true and thorough.

Yoo et al.~\citet{yoo2009comparison} instructed a group of tourism marketing students to write a hotel review from the perspective of a hotel manager. They gathered 40 truthful and 42 deceptive hotel reviews and found that truthful and deceptive reviews have different lexical complexity. Ott et al.~\citet{ott2011finding} created a much larger dataset of 800 opinions by crowdsourcing\footnote{They used Amazon Mechanical Turk} the job of writing fraudulent reviews for existing businesses. They combined work from psychology and computational linguistics to develop and compare three\footnote{Genre identification, psycholinguistic deception detection, and text categorization.} approaches for detecting deceptive opinion spam. On a similar dataset, Feng et al.~\citet{feng2012syntactic} trained Support Vector Machine model based on syntactic stylometry features for deception detection. Li et al.~\citet{li2014towards} also combined ground truth dataset created by Ott et al.~\citet{ott2011finding} with their employee (domain-expert) generated deceptive reviews to build a feature-based additive model for exploring the general rule for deceptive opinion spam detection. Rahman et al.~\citep{rahman2014turning} developed a system to detect venues that are targets of deceptive opinions. Although, this easies the identification of deceptive reviews considerable effort is still involved in identifying the actual deceptive reviews. In almost all these works, the size of the dataset limits the proposed model to reach its real capacity.

To alleviate these issues with the ground truth, we use a Generative adversarial network, which is more an unsupervised learning method rather than supervised. We start with an existing dataset and use the generator model to create necessary reviews to strengthen the classifier (discriminator). 
%


\section{Future work}
Contrary to the popular belief that supervised learning techniques are superior to unsupervised techniques, the accuracy of \sys{}, a semi-supervised learning technique is comparable to the state-of-the-art supervised techniques on the same dataset. We believe that this is a preliminary step which we plan to extend by trying different architectures like Conditional GAN~\cite{DBLP:journals/corr/MirzaO14} and better hyper-tuning.

\section{Conclusion}
In this paper, we propose \sys{}, a technique to detect deceptive reviews using Generative Adversarial Networks (GAN). To the best of our knowledge, this is the first work to leverage GANs and semi-supervised learning methods to identify deceptive reviews. Our evaluation using a dataset of 800 reviews from 20 Chicago hotels of TripAdvisor shows that \sys{} with an accuracy of~\accnosim{} performed on par with the state-of-the-art models. We believe that~\sys{} demonstrates a good first step towards using GAN for text classification tasks, specifically those requiring very large ground truth datasets. 
\section*{Acknowledgements}
We would like to thank the anonymous reviewers for their valuable comments. 
This material is based on research sponsored by the Office of Naval
Research under grant numbers N00014-15-1-2948, N00014-17-1-2011 and by DARPA under
agreement number FA8750-15-2-0084. The U.S. Government is authorized to
reproduce and distribute reprints for Governmental purposes
notwithstanding any copyright notation thereon. This work is also sponsored by a gift from Google's Anti-Abuse group. The views and conclusions contained herein are those of the authors and
should not be interpreted as necessarily representing the official
policies or endorsements, either expressed or implied, of DARPA or the
U.S. Government.
\newpage
\small
\bibliographystyle{sp2017_template/bibtex/IEEEtran}
\bibliography{refs}

\begin{thebibliography}{10}
\providecommand{\url}[1]{#1}
\csname url@samestyle\endcsname
\providecommand{\newblock}{\relax}
\providecommand{\bibinfo}[2]{#2}
\providecommand{\BIBentrySTDinterwordspacing}{\spaceskip=0pt\relax}
\providecommand{\BIBentryALTinterwordstretchfactor}{4}
\providecommand{\BIBentryALTinterwordspacing}{\spaceskip=\fontdimen2\font plus
\BIBentryALTinterwordstretchfactor\fontdimen3\font minus
  \fontdimen4\font\relax}
\providecommand{\BIBforeignlanguage}[2]{{%
\expandafter\ifx\csname l@#1\endcsname\relax
\typeout{** WARNING: IEEEtran.bst: No hyphenation pattern has been}%
\typeout{** loaded for the language `#1'. Using the pattern for}%
\typeout{** the default language instead.}%
\else
\language=\csname l@#1\endcsname
\fi
#2}}
\providecommand{\BIBdecl}{\relax}
\BIBdecl

\bibitem{Mintel2015}
M.~marketing~research company, ``Seven in 10 americans seek out opinion before
  making purchases,''
  \url{http://www.mintel.com/press-centre/social-and-lifestyle/seven-in-10-americans-seek-out-opinions-before-making-purchases},
  2015.

\bibitem{anderson2012learning}
M.~Anderson and J.~Magruder, ``Learning from the crowd: Regression
  discontinuity estimates of the effects of an online review database,''
  \emph{The Economic Journal}, vol. 122, no. 563, pp. 957--989, 2012.

\bibitem{Jindal:2008}
\BIBentryALTinterwordspacing
N.~Jindal and B.~Liu, ``Opinion spam and analysis,'' in \emph{Proceedings of
  the 2008 International Conference on Web Search and Data Mining}, ser. WSDM
  '08.\hskip 1em plus 0.5em minus 0.4em\relax New York, NY, USA: ACM, 2008, pp.
  219--230. [Online]. Available:
  \url{http://doi.acm.org/10.1145/1341531.1341560}
\BIBentrySTDinterwordspacing

\bibitem{ott2011finding}
M.~Ott, Y.~Choi, C.~Cardie, and J.~T. Hancock, ``Finding deceptive opinion spam
  by any stretch of the imagination,'' in \emph{Proceedings of the 49th Annual
  Meeting of the Association for Computational Linguistics: Human Language
  Technologies-Volume 1}.\hskip 1em plus 0.5em minus 0.4em\relax Association
  for Computational Linguistics, 2011, pp. 309--319.

\bibitem{yelp-fake-per}
B.~Technology, ``Yelp admits a quarter of submitted reviews could be fake,''
  September 2013, \url{http://www.bbc.com/news/technology-24299742}.

\bibitem{luca2016fake}
M.~Luca and G.~Zervas, ``Fake it till you make it: Reputation, competition, and
  yelp review fraud,'' \emph{Management Science}, 2016.

\bibitem{socher2011semi}
R.~Socher, J.~Pennington, E.~H. Huang, A.~Y. Ng, and C.~D. Manning,
  ``Semi-supervised recursive autoencoders for predicting sentiment
  distributions,'' in \emph{Proceedings of the conference on empirical methods
  in natural language processing}.\hskip 1em plus 0.5em minus 0.4em\relax
  Association for Computational Linguistics, 2011, pp. 151--161.

\bibitem{socher2011dynamic}
R.~Socher, E.~H. Huang, J.~Pennin, C.~D. Manning, and A.~Y. Ng, ``Dynamic
  pooling and unfolding recursive autoencoders for paraphrase detection,'' in
  \emph{Advances in neural information processing systems}, 2011, pp. 801--809.

\bibitem{socher2013recursive}
R.~Socher, A.~Perelygin, J.~Y. Wu, J.~Chuang, C.~D. Manning, A.~Y. Ng, C.~Potts
  \emph{et~al.}, ``Recursive deep models for semantic compositionality over a
  sentiment treebank,'' in \emph{Proceedings of the conference on empirical
  methods in natural language processing (EMNLP)}, vol. 1631, 2013, p. 1642.

\bibitem{elman1990finding}
J.~L. Elman, ``Finding structure in time,'' \emph{Cognitive science}, vol.~14,
  no.~2, pp. 179--211, 1990.

\bibitem{lai2015recurrent}
S.~Lai, L.~Xu, K.~Liu, and J.~Zhao, ``Recurrent convolutional neural networks
  for text classification.'' in \emph{AAAI}, vol. 333, 2015, pp. 2267--2273.

\bibitem{kim2014convolutional}
Y.~Kim, ``Convolutional neural networks for sentence classification,''
  \emph{arXiv preprint arXiv:1408.5882}, 2014.

\bibitem{zhang2015character}
X.~Zhang, J.~Zhao, and Y.~LeCun, ``Character-level convolutional networks for
  text classification,'' in \emph{Advances in neural information processing
  systems}, 2015, pp. 649--657.

\bibitem{goodfellow2014generative}
I.~Goodfellow, J.~Pouget-Abadie, M.~Mirza, B.~Xu, D.~Warde-Farley, S.~Ozair,
  A.~Courville, and Y.~Bengio, ``Generative adversarial nets,'' in
  \emph{Advances in neural information processing systems}, 2014, pp.
  2672--2680.

\bibitem{radford2015unsupervised}
A.~Radford, L.~Metz, and S.~Chintala, ``Unsupervised representation learning
  with deep convolutional generative adversarial networks,'' \emph{arXiv
  preprint arXiv:1511.06434}, 2015.

\bibitem{ehsani2017segan}
K.~Ehsani, R.~Mottaghi, and A.~Farhadi, ``Segan: Segmenting and generating the
  invisible,'' \emph{arXiv preprint arXiv:1703.10239}, 2017.

\bibitem{denton2015deep}
E.~L. Denton, S.~Chintala, R.~Fergus \emph{et~al.}, ``Deep generative image
  models using a laplacian pyramid of adversarial networks,'' in \emph{Advances
  in neural information processing systems}, 2015, pp. 1486--1494.

\bibitem{li2014towards}
J.~Li, M.~Ott, C.~Cardie, and E.~H. Hovy, ``Towards a general rule for
  identifying deceptive opinion spam.'' in \emph{ACL (1)}.\hskip 1em plus 0.5em
  minus 0.4em\relax Citeseer, 2014, pp. 1566--1576.

\bibitem{yoo2009comparison}
K.-H. Yoo and U.~Gretzel, ``Comparison of deceptive and truthful travel
  reviews,'' \emph{Information and communication technologies in tourism 2009},
  pp. 37--47, 2009.

\bibitem{metz2016unrolled}
L.~Metz, B.~Poole, D.~Pfau, and J.~Sohl-Dickstein, ``Unrolled generative
  adversarial networks,'' \emph{arXiv preprint arXiv:1611.02163}, 2016.

\bibitem{yu2017seqgan}
L.~Yu, W.~Zhang, J.~Wang, and Y.~Yu, ``Seqgan: sequence generative adversarial
  nets with policy gradient,'' in \emph{Thirty-First AAAI Conference on
  Artificial Intelligence}, 2017.

\bibitem{arjovsky2017wasserstein}
M.~Arjovsky, S.~Chintala, and L.~Bottou, ``Wasserstein gan,'' \emph{arXiv
  preprint arXiv:1701.07875}, 2017.

\bibitem{gulrajani2017improved}
I.~Gulrajani, F.~Ahmed, M.~Arjovsky, V.~Dumoulin, and A.~Courville, ``Improved
  training of wasserstein gans,'' \emph{arXiv preprint arXiv:1704.00028}, 2017.

\bibitem{goodfellow2016nips}
I.~Goodfellow, ``Nips 2016 tutorial: Generative adversarial networks,''
  \emph{arXiv preprint arXiv:1701.00160}, 2016.

\bibitem{silver2016mastering}
D.~Silver, A.~Huang, C.~J. Maddison, A.~Guez, L.~Sifre, G.~Van Den~Driessche,
  J.~Schrittwieser, I.~Antonoglou, V.~Panneershelvam, M.~Lanctot \emph{et~al.},
  ``Mastering the game of go with deep neural networks and tree search,''
  \emph{Nature}, vol. 529, no. 7587, pp. 484--489, 2016.

\bibitem{sutton2000policy}
R.~S. Sutton, D.~A. McAllester, S.~P. Singh, and Y.~Mansour, ``Policy gradient
  methods for reinforcement learning with function approximation,'' in
  \emph{Advances in neural information processing systems}, 2000, pp.
  1057--1063.

\bibitem{goodfellow2016deep}
I.~Goodfellow, Y.~Bengio, and A.~Courville, \emph{Deep learning}.\hskip 1em
  plus 0.5em minus 0.4em\relax MIT Press, 2016.

\bibitem{hochreiter1997long}
S.~Hochreiter and J.~Schmidhuber, ``Long short-term memory,'' \emph{Neural
  computation}, vol.~9, no.~8, pp. 1735--1780, 1997.

\bibitem{zhang2015text}
X.~Zhang and Y.~LeCun, ``Text understanding from scratch,'' \emph{arXiv
  preprint arXiv:1502.01710}, 2015.

\bibitem{wang2017liar}
W.~Y. Wang, ``" liar, liar pants on fire": A new benchmark dataset for fake
  news detection,'' \emph{arXiv preprint arXiv:1705.00648}, 2017.

\bibitem{srivastava2015highway}
R.~K. Srivastava, K.~Greff, and J.~Schmidhuber, ``Highway networks,''
  \emph{arXiv preprint arXiv:1505.00387}, 2015.

\bibitem{abadi2016tensorflow}
M.~Abadi, A.~Agarwal, P.~Barham, E.~Brevdo, Z.~Chen, C.~Citro, G.~S. Corrado,
  A.~Davis, J.~Dean, M.~Devin \emph{et~al.}, ``Tensorflow: Large-scale machine
  learning on heterogeneous distributed systems,'' \emph{arXiv preprint
  arXiv:1603.04467}, 2016.

\bibitem{feng2012syntactic}
S.~Feng, R.~Banerjee, and Y.~Choi, ``Syntactic stylometry for deception
  detection,'' in \emph{Proceedings of the 50th Annual Meeting of the
  Association for Computational Linguistics: Short Papers-Volume 2}.\hskip 1em
  plus 0.5em minus 0.4em\relax Association for Computational Linguistics, 2012,
  pp. 171--175.

\bibitem{pennington2014glove}
\BIBentryALTinterwordspacing
J.~Pennington, R.~Socher, and C.~D. Manning, ``Glove: Global vectors for word
  representation,'' in \emph{Empirical Methods in Natural Language Processing
  (EMNLP)}, 2014, pp. 1532--1543. [Online]. Available:
  \url{http://www.aclweb.org/anthology/D14-1162}
\BIBentrySTDinterwordspacing

\bibitem{drucker1999support}
H.~Drucker, D.~Wu, and V.~N. Vapnik, ``Support vector machines for spam
  categorization,'' \emph{IEEE Transactions on Neural networks}, vol.~10,
  no.~5, pp. 1048--1054, 1999.

\bibitem{gyongyi2004combating}
Z.~Gy{\"o}ngyi, H.~Garcia-Molina, and J.~Pedersen, ``Combating web spam with
  trustrank,'' in \emph{Proceedings of the Thirtieth international conference
  on Very large data bases-Volume 30}.\hskip 1em plus 0.5em minus 0.4em\relax
  VLDB Endowment, 2004, pp. 576--587.

\bibitem{ntoulas2006detecting}
A.~Ntoulas, M.~Najork, M.~Manasse, and D.~Fetterly, ``Detecting spam web pages
  through content analysis,'' in \emph{Proceedings of the 15th international
  conference on World Wide Web}.\hskip 1em plus 0.5em minus 0.4em\relax ACM,
  2006, pp. 83--92.

\bibitem{gyongyi2005web}
Z.~Gyongyi and H.~Garcia-Molina, ``Web spam taxonomy,'' in \emph{First
  international workshop on adversarial information retrieval on the web
  (AIRWeb 2005)}, 2005.

\bibitem{wu2010distortion}
G.~Wu, D.~Greene, B.~Smyth, and P.~Cunningham, ``Distortion as a validation
  criterion in the identification of suspicious reviews,'' in \emph{Proceedings
  of the First Workshop on Social Media Analytics}.\hskip 1em plus 0.5em minus
  0.4em\relax ACM, 2010, pp. 10--13.

\bibitem{rahman2014turning}
M.~Rahman, B.~Carbunar, J.~Ballesteros, G.~Burri, D.~Horng \emph{et~al.},
  ``Turning the tide: Curbing deceptive yelp behaviors.'' in \emph{SDM}.\hskip
  1em plus 0.5em minus 0.4em\relax SIAM, 2014, pp. 244--252.

\bibitem{DBLP:journals/corr/MirzaO14}
\BIBentryALTinterwordspacing
M.~Mirza and S.~Osindero, ``Conditional generative adversarial nets,''
  \emph{CoRR}, vol. abs/1411.1784, 2014. [Online]. Available:
  \url{http://arxiv.org/abs/1411.1784}
\BIBentrySTDinterwordspacing

\end{thebibliography}

\end{document}